\documentclass[conference, letterpaper]{IEEEtran}

\usepackage[T1]{fontenc}
\usepackage[latin1]{inputenc}
\usepackage{color}
\usepackage{psfrag}
\usepackage[dvips]{graphicx}
\usepackage{tikz}
\usepackage{pgfplots}
\pgfplotsset{compat=1.12}
\usepackage{enumerate}
\usepackage{rotating}
\usepackage{srcltx}
\usepackage{booktabs}
\usepackage{multicol}
\usepackage{enumitem}
\usepackage{textcomp}
\usetikzlibrary{shapes,arrows}
\usepackage{array,multirow,graphicx}
\usepackage{multirow}
\usetikzlibrary{arrows,decorations.pathmorphing,backgrounds,positioning,fit,matrix}

\usepackage[noadjust]{cite}
\usepackage[cmex10]{amsmath}
\usepackage{amsfonts}
\usepackage{amssymb}
\usepackage{bbm}
\usepackage{ntheorem}
\usepackage{dsfont}    
\usepackage{mathtools} 
\usepackage{stackengine}
\usepackage{enumitem}
\usepackage{xcolor}

\usepackage{algorithmic}
\usepackage{array}
\usepackage{graphicx}
\usepackage{caption}
\usepackage{subcaption}

\theoremseparator{.}
\newtheorem{corollary}{Corollary}
\newtheorem{prop}{Proposition}
\newtheorem{lem}{Lemma}
\newtheorem{theorem}{Theorem}
\theorembodyfont{\upshape}
\theoremseparator{.}
\newtheorem{definition}{Definition}
\newtheorem{remark}{Remark}
\newtheorem*{example*}{Example}

\newcommand{\eu}{\mathrm{e}}

\newcommand{\E}{\mathbb{E}}

\newcommand{\supp}{{\mathsf{supp}}}

\allowdisplaybreaks

\long\def\symbolfootnote[#1]#2{\begingroup%
	\def\thefootnote{\fnsymbol{footnote}}\footnote[#1]{#2}\endgroup} 
\title{Estimation Error: Distribution and Pointwise Limits }
\IEEEoverridecommandlockouts    
 \author{
 \IEEEauthorblockN{ Luca Barletta$^{*}$, Alex Dytso$^{\dagger}$, and Shlomo Shamai (Shitz)$^{**}$}
 $^{*}$ Politecnico di Milano, Milano, 20133, Italy. Email: luca.barletta@polimi.it \\
 $^{\dagger}$ Qualcomm Flarion Technologies, Bridgewater,  NJ 08807, USA.
 Email: odytso2@gmail.com \\
 $^{**}$  Technion -- Israel Institute of Technology, Haifa 32000, Israel. E-mail: sshlomo@ee.technion.ac.il}

\begin{document}

\maketitle

\begin{abstract}
     In this paper, we examine the distribution and convergence properties of the estimation error $W = X - \hat{X}(Y)$, where $\hat{X}(Y)$ is the Bayesian estimator of a random variable $X$ from a noisy observation $Y = X +\sigma Z$ where $\sigma$ is the parameter indicating the strength of noise $Z$. Using the conditional expectation framework (that is, $\hat{X}(Y)$ is the conditional mean), we define the normalized error $\mathcal{E}_\sigma = \frac{W}{\sigma}$ and explore its properties. 
     
    Specifically, in the first part of the paper, we characterize the probability density function of $W$ and $\mathcal{E}_\sigma$.  Along the way, we also find conditions for the existence of the inverse functions for the conditional expectations. In the second part, we study pointwise (i.e., almost sure)  convergence of $\mathcal{E}_\sigma$ as $\sigma \to 0$ under various assumptions about the noise and the underlying distributions. Our results extend some of the previous limits of $\mathcal{E}_\sigma$ as $\sigma \to 0$ studied under the $L^2$ convergence, known as the \emph{MMSE dimension}, to the pointwise case. 

\end{abstract}

\section{Introduction}
Consider a setting where we seek to estimate a random variable $X$ from a noisy observation $Y$.  One accepted way to compute the error is to assess the difference\footnote{Other errors are also common; see for example \cite{BregmanCE}.} between \(X\) and the estimate \(\hat{X}(Y)\) (e.g., the Bayesian estimator):
\begin{equation}
W = X - \hat{X}(Y).
\end{equation}
The `quality' of the estimate is then typically evaluated by computing some absolute \(p\)-th moment of \(W\), of which the second moment, known as the mean squared error, is by far the most widely used. While the moments of \(W\) have received attention in the literature, the distribution of \(W\) itself has not been thoroughly investigated. The goal of this work is to start making progress in understanding other properties of \(W\), such as its distribution and the rate of convergence of \(W\) as some noise parameter (e.g., noise variance) goes to zero, or when some signal-to-noise ratio parameter goes to infinity. Practically, understanding the full distribution of the estimation error 
$W$ provides deeper insight into the reliability and risk of extreme errors in practical systems beyond what average measures like mean squared error can reveal.

Specifically, in this paper, we focus on scenarios where the noise is modeled as additive, that is:
\begin{equation}
Y = X + \sigma Z
\end{equation}
where \(X\) and \(Z\) are independent random variables and \(\sigma > 0\) is a noise parameter. Moreover, as the estimator, we take the conditional expectation:
\begin{equation}
\hat{X}(Y) = \mathbb{E}[X|Y]
\end{equation}
which is an optimal Bayesian estimator under the second moment criterion. We also define the normalized error as:
\begin{align}
\mathcal{E}_\sigma 
= \frac{X - \mathbb{E}[X|Y]}{\sigma}
= \frac{W}{\sigma}
\end{align}
and seek to study the almost sure (a.s.) convergence of \(\mathcal{E}_\sigma\) as $\sigma \to 0$.

\subsection{Definitions and Notation}
Throughout the paper, the probability space $(\Omega, \mathcal{F}, \mathbb{P})$ is fixed.  The density of a random variable  $X$ whose distribution is absolutely continuous with respect to Lebesgue measure is denoted by $f_X$. $\phi_{\sigma}$ denotes the density of Gaussian with  zero mean and  variance $\sigma^2$.   If measure $\mu  $ is absolutely continuous with respect to $\lambda$, then it is denoted as $\mu \ll \lambda$. Similarly, $\mu \perp \nu$ denotes that $\mu$ and $\nu$ are mutually singular. $L^1$ denotes the collection of all random variables defined on $(\Omega, \mathcal{F}, \mathbb{P})$ with finite first absolute moment.   The conditional expectation of $X$ given $Y$ is defined as
\begin{equation}
\mathbb{E}[X|Y =y] = \int x {\rm d} P_{X|Y=y}(x)
\end{equation}
where $P_{X|Y=y}(x)$ is the conditional distribution. All logarithms are base $\eu$.

\subsection{Paper Outline and Contributions}

\begin{enumerate}
\item Section~\ref{sec:Distr_CE} studies the distribution of $W$ and $\mathcal{E}_\sigma$ and shows:
\begin{itemize}
\item Section~\ref{sec:Inverse_OF_CE}, Proposition~\ref{prop:inverse-existance}, derives sufficient conditions for  $y \mapsto \mathbb{E}[X|Y=y]$ to have a functional inverse, which is needed to study the distribution of $W$. In particular, it is shown that if  certain conditions  hold, most notably $Z$ has a log-concave distribution, then the inverse exists.
\item Section~\ref{sec:pdf_distribution}, Theorem~\ref{thm:distribution_of_errors}, characterizes the density function of $W$ and evaluates it for some examples. 
\end{itemize}
\item Section~\ref{sec:limit_of_error} studies the a.s. convergence of $\mathcal{E}_\sigma$ under various assumptions on the distribution of $X$ and distribution of $Z$. 
\end{enumerate}
The rest of this section is dedicated to literature review. Finally, some of the proofs are omitted due to length constraints and can be found in the Technical Appendix.

\subsection{Related Literature}

Conditional expectation plays an important role in our discussion.  Monotonicity of  conditional expectation under rather general settings has been shown in  \cite{Witsenhausen1969, PiotrBolesawNowak2013}. The derivatives of the conditional expectation under additive and exponential family models have been considered in \cite{hatsell1971some,condMeanDer,CEderivativeExpo} where the derivative were shown to be proportional to conditional cumulants.  The probability distribution for the Gaussian noise has been previously found in \cite{distCEITW}. In \cite{venkat2012pointwise},  in the context of Gaussian noise, $W^2$
has been related to the information density via the so-called pointwise I-MMSE relationship. Other related properties of $W$ and the  conditional probability $P_{X|Y}$ have been studied in \cite{guo2005mutual, guo2011estimation,guo2005additive,palomar2007representation,rugini2016equivalence}; see \cite{guo2013interplay} for a comprehensive survey of results and applications.

In \cite{wu2011mmse}, authors considered   convergence of the second moment of  $\mathcal{E}_\sigma$ (i.e., $ \mathbb{E} [(\mathcal{E}_\sigma)^2]$), which was termed \emph{MMSE dimension}, and provided a fairly complete characterization of the limit.  For example, under suitable regularity conditions of the noise distribution $Z$, if the distribution $P_X = \alpha P_{X_c} + (1-\alpha) P_{X_D}, \, \alpha \in [0,1]$ where $P_{X_c}$ is absolutely continuous distribution with respect to Lebesgue measure and $P_{X_D}$ is a discrete distribution,  then $\lim_{\sigma \to 0} \mathbb{E} [(\mathcal{E}_\sigma)^2] = \alpha {\rm Var}(Z)$.
We will provide the pointwise version of this result. 
Some of the limit theorems shown in \cite{wu2011mmse} would be useful in our analysis too.  To find limits, we also borrow techniques from~\cite{doob1960relative}.

\section{On the Distribution of the Estimation Error }
\label{sec:Distr_CE}

In this section, we derive the distribution of $X-\E[X|Y]$. To fully derive the result, we need to characterize conditions under which the functional inverse $y \mapsto \E[X|Y=y]$ exists. In addition, we need conditions for this inverse to be differentiable.

\subsection{On the Inverse Function of the Conditional Expectation}
\label{sec:Inverse_OF_CE}

In this section, we study the inverse function of the conditional expectation.  For  ease of notation, we let $\hat{X}(y)=\E[X|Y=y]$, and let $\hat{X}^{-1}$ denote the functional inverse provided that it exists.  The next proposition gives a sufficient condition for the existence of the inverse.  

\begin{prop} \label{prop:inverse-existance}
Suppose that 
\begin{itemize}
\item $X$ is non-degenerate and $X \in L^1$; and
\item the noise density function can be written as 
\begin{equation}
f_Z(z) = \left \{ \begin{array}{cc}
\eu^{\psi(z)} & z \in I\\
0 & \text{ o.w.}
\end{array} \right.
\end{equation}
where $I$ is an open interval (possibly infinite) and $z \mapsto \psi(z)$ is strictly concave on $I$\footnote{This property is known as log-concavity if $I =\mathbb{R}$.}; and 
\item    there exists a function $\theta(X) \in L^1$ such that for all $y \in \mathbb{R}$
\begin{equation}
| X \psi'(y-X) \exp( \psi(y-X)) | \le \theta(X) \text{ a.s. } \label{eq:existance_der_cond}
\end{equation}
\end{itemize}
Then,  $\hat{X}^{-1}$ exists and is differentiable.  

\end{prop}
\begin{IEEEproof}
Under the first two assumptions, by slightly modifying the argument of \cite[Prop~.1]{Witsenhausen1969} -- the modification involves assuming that $\psi$ is strictly concave instead of concave -- the conditional expectation $y \mapsto \hat{X}(y)$ is an increasing function  instead of just non-decreasing. Since increasing functions have proper functional inverses, the first part of the conclusion follows. 

Next, note that 
\begin{equation}
 \hat{X}(y) = \frac{\E \left[ X  \exp(\psi(y-X) )\right]}{\E \left[  \exp(\psi(y-X) )\right]} 
\end{equation}
 which, by employing Leibniz integral rule to numerator and denominator, is differentiable provided that 
\begin{align}
    | X \psi'(y-X) \exp( \psi(y-X)) | &\le \theta_1(X)  \text{ a.s. }\label{eq;top_diff}\\
    | \psi'(y-X) \exp( \psi(y-X)) | &\le \theta_2(X)  \text{ a.s. },\label{eq;bottom_diff}
\end{align}
for some $L^1$ functions $\theta_1$ and $\theta_2$. The proof is concluded by noting that \eqref{eq;top_diff} subsumes \eqref{eq;bottom_diff} resulting in condition \eqref{eq:existance_der_cond}. 
\end{IEEEproof}

The condition of Proposition \ref{prop:inverse-existance} is only sufficient. In particular, in the case where $Z$ is Gaussian, the existence of the inverse follows from the fact that the derivative of $\hat{X}(y)$ for non-degenerate $X$ is strictly positive since  \cite{condMeanDer,hatsell1971some}
\begin{equation}
  \sigma^2 \frac{{\rm d}}{{\rm d} y} \E[X|Y=y]=  \mathsf{Var}(X|Y=y),  \quad y \in \mathbb{R},  \label{eq:HN_identity}
\end{equation}
which holds for all distributions on $X$.

\begin{example*} Suppose that $X$ and $Z$  are two independent standard Gaussian random variables.  Then, the conditional expectation is given by 
\begin{equation}
\hat{X}(y)=   \frac{1}{1+\sigma^2} y, \quad y \in \mathbb{R}, 
\end{equation}
and the inverse is given by 
\begin{equation}
 \hat{X}^{-1}(x)= (1+\sigma^2) x,  \quad    x\in \mathbb{R}. 
 \end{equation} 
\end{example*}

\begin{example*} Suppose that $X$  is distributed according to $P_X(1)=p=1-P_X(-1)$.   Then, the conditional expectation is given by
\begin{equation}
\E[X|Y=y]=  \tanh \left(  \frac{y}{\sigma^2} + \frac{1}{2} \log \left( \frac{p}{1-p} \right)   \right),\, y\in \mathbb{R},    \label{eq:tahn_p_classifier}
\end{equation} 
and the inverse is given by 
\begin{equation}
\hat{X}^{-1}(x)=  \frac{\sigma^2}{2} \log \left( \frac{1+x}{1-x} \frac{1-p}{p} \right), \quad x  \in  (-1,1).  \label{eq:tahn_p_classifier_inverse} 
\end{equation} 
\end{example*}

\subsection{Characterization of the Estimation Error Distribution}
\label{sec:pdf_distribution}

The next theorem provides an expression for the probability density function (pdf) of the estimation error $X-g(Y)$ where $g$ is  some estimator of $X$. 
\begin{theorem}\label{thm:distribution_of_errors} Let $W=X-g(Y)$ where $g$ has a well-defined functional inverse and where  $\mathcal{R}_g$ denotes the range of the function $g$. Then, for $ w\in \mathbb{R}$
\begin{align}
f_W(w)  
&= \frac{1}{\sigma} \E \bigg[ f_Z\left( \frac{ g^{-1}(X-w)-X}{\sigma} \right)   \left|  \frac{{\rm d} g^{-1}(X-w) }{ {\rm d} w} \right|  \nonumber\\
& \qquad\qquad \cdot\mathbbm{1}_{\mathcal{R}_g }   \left( w-X \right)   \bigg]. \label{eq:ExpressionForTHeError_general_g} 
\end{align}
Consequently, if  $g(Y)= \hat{X}(Y)$, for $ w\in \mathbb{R}$
\begin{align}
 f_W(w) &=\frac{1}{\sigma} \E   \bigg[    f_Z  \left(  \frac{ \hat{X}^{-1}  \left(X-w \right)-X }{\sigma}  \right)   \left|    \frac{{\rm d}  \hat{X}^{-1}(X-w) }{ {\rm d} w}  \right| \nonumber \\
&\qquad \qquad \cdot \mathbbm{1}_{ \mathcal{R}_{\hat{X}}  }   \left( w-X \right)     \bigg]. \label{eq:Distribution_Error_MMSE}
\end{align} 

\end{theorem}
\begin{IEEEproof}
Let $W=X-g(X+N)=X+U$, where $N=\sigma Z$; then,  by using the formula for  the sum of correlated variables, we have that 
\begin{equation}
f_W(w)=\E \left[  f_{U|X}(w-X|X)  \right].  \label{eq:GeneralConvolutionExpression}
\end{equation}
To characterize $f_{U|X}(t|x)$, note that given $X=x$ 
\begin{equation}
U=-g(x+N) = -g(N_x) 
\end{equation}
where $N_x=N+x$. 
Therefore,
\begin{align}
f_{U|X}(t|x)&= f_{ - g(N_x) }(t)\\
&= f_{N_x}(g^{-1}(-t)) \left| \frac{ {\rm d} }{ {\rm d} t } g^{-1}(-t) \right|     \mathbbm{1}_{ \mathcal{R}_{g}  }(t)  \label{eq:Change_Variable_ErrorFormula}\\
&=  f_{\sigma Z} \left(   g^{-1}(-t)-x \right)  \left| \frac{ {\rm d}}{{\rm d} t } g^{-1}(-t) \right|  \mathbbm{1}_{ \mathcal{R}_{g}  }(t) \\
&=  \frac{1}{\sigma} f_{ Z} \left(  \frac{ g^{-1}(-t)-x}{\sigma} \right)  \left| \frac{ {\rm d}}{{\rm d} t } g^{-1}(-t) \right|  \mathbbm{1}_{ \mathcal{R}_{g}  }(t)
, \label{eq:PDF_U_given_X_err}
\end{align} 
where in \eqref{eq:Change_Variable_ErrorFormula} we have used a change of variable formula.  Inserting \eqref{eq:PDF_U_given_X_err} into \eqref{eq:GeneralConvolutionExpression} concludes the proof of \eqref{eq:ExpressionForTHeError_general_g}. 
\end{IEEEproof} 

\begin{remark}
Both of the expressions in Theorem~\ref{thm:distribution_of_errors} can be further simplified or rewritten.  
In particular, the expression in \eqref{eq:ExpressionForTHeError_general_g} can be rewritten by using  $\frac{{\rm d}}{ {\rm d} t} g^{-1}(t)=\frac{1}{g'( g^{-1}(t) ) }$. For example, when $Z$ is standard Gaussian,  by using \eqref{eq:HN_identity} we have that 
\begin{align}
   \left|  \hspace{-0.01cm}  \frac{{\rm d}  \hat{X}^{-1} \hspace{-0.02cm}(w) }{ {\rm d} w}  \hspace{-0.01cm}\right| = \frac{\sigma^2}{\mathsf{Var}\left( X|Y= \hat{X}^{-1} \hspace{-0.02cm}(w) \right)}, 
\end{align}   
which leads to: for   $w\in \mathbb{R}$
\begin{equation}
f_W(w) 
=  \sigma^2 \hspace{-0.01cm}\E  \hspace{-0.05cm} \left[    \hspace{-0.01cm}  \frac{ \hspace{-0.05cm} \phi_{\sigma} \hspace{-0.05cm} \left( \hspace{-0.02cm} \hat{X}^{-1} \hspace{-0.05cm} \left(X-w \right)-X \hspace{-0.02cm}  \right) }{ \mathsf{Var}\left( X|Y= \hat{X}^{-1} \hspace{-0.02cm}(X-w) \right)}  \mathbbm{1}_{ \mathcal{R}_{\hat{X}}   } \left(w-X \right)  \hspace{-0.05cm} \right]  \hspace{-0.02cm}.\label{eq:Estimation_Error_pdf}
\end{equation}

\end{remark}

We note that if one seeks to find the distribution of $\mathcal{E}_\sigma=\frac{W}{\sigma}$ instead of $W$, then the  transformation $f_{\mathcal{E}_\sigma}(w) = \sigma f_{W}(\sigma w)$, $w \in \mathbb{R}$ can be used.

\begin{figure}[t]
\vspace{1ex}
\centering
\resizebox{0.7\columnwidth}{!}{\input{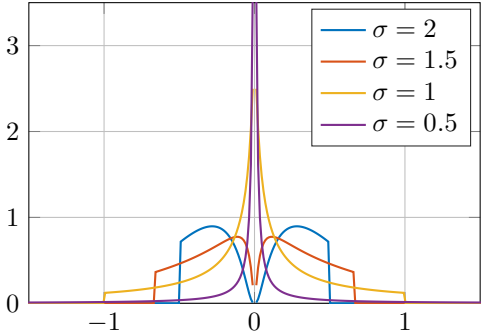}}
\vspace{1ex}
\caption{Density of $\mathcal{E}_\sigma$ in \eqref{eq:Binary_Symmetric}.}
\label{fig:PdfOfError}
\end{figure}

\begin{example*}Suppose that $X$  is distributed according to $P_X(1)=1-P_X(-1)=p \in (0,1)$. The conditional expectation of the random variable is given in \eqref{eq:tahn_p_classifier}, and the inverse is given in \eqref{eq:tahn_p_classifier_inverse}.  The derivative of the inverse is given by $\frac{{\rm d}}{ {\rm d} x} \hat{X}^{-1}(x)=  \frac{\sigma^2}{1-x^2}, x  \in  \mathcal{D}_{\hat{X}} $ where $\mathcal{D}_{\hat{X}} =(-1,1)$. Therefore, by using Theorem~\ref{thm:distribution_of_errors},  the distribution of the error is given by 
\vspace{-0.5em}
\begin{align}
 &f_{\mathcal{E}_\sigma}(w) 
=   \phi_{\sigma} \left(   \frac{\sigma^2}{2} \log \left( \frac{2-\sigma w}{\sigma w} \frac{1-p}{p} \right) -1 \right)   \frac{\sigma^3 p}{1-(1-\sigma w)^2} \nonumber\\
&\quad \cdot \mathbbm{1}_{ (0,2) }(\sigma w)  \notag \\
&\quad +\phi_{\sigma} \left(   \frac{\sigma^2}{2} \log \left( \frac{-\sigma w}{2+\sigma w} \frac{1-p}{p} \right) +1 \right)   \frac{\sigma^3 (1-p)}{1-(1+\sigma w)^2} \nonumber\\
&\quad \cdot \mathbbm{1}_{  (-2,0) }(\sigma w) , \quad w\in \mathbb{R},  \label{eq:pdf_erro_binary}
\end{align}
\vspace{-0.5em}
which in the case of $p=1/2$ reduces to 
\begin{align}
 f_{\mathcal{E}_\sigma}(w) &=  \frac{1}{2}  \phi_{\sigma} \left(  \frac{\sigma^2}{2} \log \left( \frac{2-|\sigma w|}{|\sigma w|}  \right)-1 \right)    \frac{\sigma^3}{1-(1-|\sigma w|)^2} \nonumber\\
 &\quad \cdot \mathbbm{1}_{  (-2,2) }(\sigma w), \quad w\in \mathbb{R}. \label{eq:Binary_Symmetric}
\end{align}


Figure~\ref{fig:PdfOfError} displays the density in \eqref{eq:Binary_Symmetric}. 
\end{example*} 

\section{On 
 $\lim_{\sigma \to 0} \mathcal{E}_\sigma = \mathcal{E}_0$ }
 \label{sec:limit_of_error}

 In this section, we find the distribution of $\mathcal{E}_0$ under various assumption on the distributions of $X$ and $Z$.   Table~\ref{tab:sample_table} summarizes the results. 

\begin{table*}[h!]
\centering
\caption{The expression for $\lim_{\sigma \to 0} \mathcal{E}_\sigma = \mathcal{E}_0$ under various assumptions.}
\label{tab:sample_table}
 \begin{tabular}{|c|c|c|}
\hline
Assumption on $X$ & Assumption on $Z$ & $\lim_{\sigma \to 0} \mathcal{E}_\sigma = \mathcal{E}_0$ \\
\hline\hline
discrete distribution & $f_Z$ bounded  and $o(|z|^{-1})$ & $0$ \\
\hline
bounded and continuous density & $Z \in L^1$ & \multirow{2}{*}{$\mathbb{E}[Z]-Z$} \\
 \cline{1-2}
absolutely cont. distribution & $f_Z$ bounded  and $O(|z|^\alpha)$ for some $\alpha>2$ & \\
 \hline
 $\begin{array}{c}
X = \mathbbm{1}_{\{U=1\}}X_D + \mathbbm{1}_{\{U=2\}} X_c  \\ U \perp X_D \perp X_c  \\ P_{X_D} - \text{discrete, } P_{X_c} - \text{abs. continuous}  \\
U \in \{1,2\}
\end{array}$
& Doob's random variable (see Definition~1) &  $\mathbbm{1}_{\{U=2\}}(\mathbb{E}[Z]-Z)$ \\
\hline
\end{tabular}

\end{table*}
\subsection{Absolutely Continuous Distributions}

In this section, we assume that the distribution of $X$ has a density with respect to Lebesgue measure.

\begin{theorem}
    Suppose that 
    \begin{itemize}
        \item   the density  $f_X$ is continuous and bounded; and
        \item   $Z \in L^1$. 
    \end{itemize} 
    Then, 
    \begin{equation}
       \lim_{\sigma \to 0} {\cal E}_\sigma = \mathbb{E}[Z]-Z \qquad \text{a.s.}
    \end{equation}
\end{theorem}
\begin{IEEEproof}
First observe that 
\begin{align}
{\cal E}_\sigma &= \frac{X - \E[X|Y]}{\sigma} \\
&= \frac{X - Y + Y - \E[X|Y]}{\sigma}  \\
&= -  Z +  \E[Z|Y]. \label{eq:E_sigma_as_Z_error}
\end{align}
Next, consider for any $x,z \in \mathbb{R}$:
\begin{align}
    \lim_{\sigma \to 0} \E[Z|Y=x+\sigma z] &=\lim_{\sigma \to 0} \frac{\E[Z f_X(x+\sigma z-\sigma Z)]}{\E[f_X(x+\sigma z-\sigma Z)]} \\
    &= \frac{\E[Z \lim_{\sigma \to 0} f_X(x+\sigma z-\sigma Z)]}{\E[ \lim_{\sigma \to 0}f_X(x+\sigma z-\sigma Z)]} \label{eq:apply_DCT} \\
    &= \frac{\E[Z] f_X(x)}{f_X(x)}, \label{eq:apply_continuity} 
\end{align}
where \eqref{eq:apply_DCT} follows from boundedness of $f_X$, from $Z \in L^1(\Omega)$, and from dominated convergence theorem; \eqref{eq:apply_continuity} follows from continuity of $f_X$. Since $f_X(X)>0$ a.s., we have $\lim_{\sigma \to 0} \E[Z|Y] = \E[Z]$ a.s. which, combined with \eqref{eq:E_sigma_as_Z_error}, gives the claimed result.
\end{IEEEproof}

We now put more assumptions  on the noise $Z$ while keeping minimal assumption on $X$. 
\begin{theorem}
    Suppose that  $Z$ has a bounded density and such that for some $\alpha > 2$,
    \begin{equation}
        f_Z(z) = O(|z|^{-\alpha}), \qquad |z|\to \infty.
    \end{equation}
    Then,  for all $X$ with absolutely continuous distribution 
    \begin{equation}
       \lim_{\sigma \to 0} {\cal E}_\sigma = \mathbb{E}[Z]-Z \qquad \text{a.s.}
    \end{equation}

\end{theorem}
\begin{IEEEproof}
    Let us introduce the density of the variable $\sigma(Z-z)$:
    \begin{equation}
        K_\sigma(x) = \frac{1}{\sigma}f_Z\left(\frac{x}{\sigma} + z \right).
    \end{equation}
    Notice that
    \begin{equation}
        f_Y(x+\sigma z) = f_X \ast K_\sigma (x),
    \end{equation}
   where $*$ denotes convolution.
    Then, by using $f_X \in L^1(\mathbb{R})$, the conditions on $f_Z$, and \cite[Lemma 8]{wu2011mmse}, we can say that
    \begin{equation}\label{eq:conv_Ksigma}
        \lim_{\sigma\to 0} f_Y(x+\sigma z) = f_X(x)
    \end{equation}
    holds for Lebesgue-a.e.~$x$. Let us check that the conditions of \cite[Lemma 8]{wu2011mmse} for the function $K_\sigma$ are satisfied:
    \begin{itemize}
        \item $\int_{\mathbb{R}} K_\sigma(x) \mathrm{d}x=1$;
        \item $\sup_{x \in \mathbb{R}, \, \sigma>0} \sigma |K_\sigma(x)|=\sup_{x \in \mathbb{R}} |f_Z(x)|<\infty$ because of the boundedness condition on $f_Z$;
        \item \begin{align}
         &\left(\sup_{x \in \mathbb{R}, \, \sigma>0} \frac{|x|^{1+\eta}}{\sigma^\eta} |K_\sigma(x)|\right)^{\frac{1}{1+\eta}}    \\
         &=\sup_{u\in\mathbb{R}} |u|f_Z(u+z)^{\frac{1}{1+\eta}} \\
         &\le \sup_{u\in\mathbb{R}} |u|f_Z(u)^{\frac{1}{1+\eta}}+|z|\sup_{u\in\mathbb{R}} f_Z(u)^{\frac{1}{1+\eta}} \\
         &< \infty
        \end{align}
        because of the boundedness condition on $f_Z$ and because $f_Z(u) = O(|u|^{-2-\eta})$ for some $\eta>0$.
    \end{itemize}
    
    Next, introduce the function
    \begin{equation}
        G_\sigma(x) = \left(\frac{x}{\sigma}+z \right) \frac{1}{\sigma}f_Z\left(\frac{x}{\sigma}+z \right),
    \end{equation}
    and notice that
    \begin{equation}
        \E[Z f_X(x+\sigma z-\sigma Z)] = f_X \ast G_\sigma (x).
    \end{equation}
    Then, by using $f_X \in L^1(\mathbb{R})$, the conditions on $f_Z$, and \cite[Lemma 8]{wu2011mmse}, we can say that \newpage
    \begin{equation}\label{eq:conv_Gsigma}
        \lim_{\sigma\to 0} \E[Z f_X(x+\sigma z-\sigma Z)] = \E[Z] f_X(x)
    \end{equation}
    holds for Lebesgue-a.e.~$x$. Let us check that the conditions of \cite[Lemma 8]{wu2011mmse} for the function $G_\sigma$ are satisfied:
    \begin{itemize}
        \item $\int_{\mathbb{R}} G_\sigma(x) \mathrm{d}x=\E[Z]$;
        \item $\sup_{x \in \mathbb{R}, \, \sigma>0} \sigma |G_\sigma(x)|=\sup_{u \in \mathbb{R}} |u|f_Z(u)<\infty$ because of the boundedness condition on $f_Z$ and because $f_Z(u) = O(|u|^{-2-\eta})$ for some $\eta>0$;
        \item 
        \begin{align}
            &\left(\sup_{x \in \mathbb{R}, \, \sigma>0} \frac{|x|^{1+\eta}}{\sigma^\eta} |G_\sigma(x)|\right)^{\frac{1}{1+\eta}} \\
            &= \sup_{u\in\mathbb{R}} |u| \left|u+z\right|^{\frac{1}{1+\eta}}f_Z(u+z)^{\frac{1}{1+\eta}} \\
            &\le \sup_{u\in\mathbb{R}} |u|^{1+\frac{1}{1+\eta}}f_Z(u)^{\frac{1}{1+\eta}}+|z|\sup_{u\in\mathbb{R}} |u|^{\frac{1}{1+\eta}} f_Z(u)^{\frac{1}{1+\eta}} \nonumber\\
            &<\infty
            \end{align}
            because of the boundedness condition on $f_Z$ and because $f_Z(u) = O(|u|^{-2-\eta})$ for some $\eta>0$.
    \end{itemize}
    
    Using the results \eqref{eq:conv_Ksigma} and \eqref{eq:conv_Gsigma}, we have 
    \begin{align}
        \lim_{\sigma\to 0} \E[Z|Y= x+\sigma z] &= \lim_{\sigma \to 0} \frac{\E[Z f_X(x+\sigma z-\sigma Z)]}{\E[f_X(x+\sigma z-\sigma Z)]}\\
        &=\E[Z] \quad \text{a.s.}
    \end{align}
\end{IEEEproof}

\subsection{Discrete Distributions}
We now consider discrete distributions. The next theorem makes the statement that \emph{discrete random variables are `easy' to estimate} more concrete. 
\begin{theorem}
    If $X$ is discrete (finitely or countable infinitely valued), then
    \begin{equation}
        \lim_{\sigma\to 0} {\cal E}_\sigma = 0 \quad \text{a.s.}
    \end{equation}
    holds for all $Z$ with bounded density and such that
    \begin{equation}
        f_Z(z) = o(|z|^{-1}), \qquad |z|\to \infty.
    \end{equation}
\end{theorem}
\begin{IEEEproof}
    Start from
    \begin{equation}
        {\cal E}_\sigma  = \E[Z|Y]-Z.
    \end{equation}
    Let $p_i:=\Pr(X=x_i)$. For all $z$ and $x_i$, define the function
    \begin{equation}
        g_{z,i}(\sigma) := |z-\E[Z|Y=x_i+\sigma z]|.
    \end{equation}
    Then, for all $z \in \mathbb{R}$ and $x_i \in \supp(P_X)$ we have
    \newpage
    \begin{align}
        g_{z,i}(\sigma) &= \frac{\left|\E\left[ \frac{x_i-X}{\sigma} f_Z\left(\frac{x_i-X}{\sigma}+z \right)\right]\right|}{\sum_{j} p_j f_Z\left(\frac{x_i-x_j}{\sigma}+z \right)} \\
        &\le \frac{\E\left[ \frac{|x_i-X|}{\sigma} f_Z\left(\frac{x_i-X}{\sigma}+z \right)\right]}{p_i f_Z\left(z \right)}. \label{eq:firstBound_on_g}
    \end{align}
    Next, fix some $d > 0$ and note that
    \begin{align}
&\frac{|x_i-X|}{\sigma} f_Z\left(\frac{x_i-X}{\sigma}+z \right) \notag\\
&= \frac{|x_i-X|}{\sigma} f_Z\left(\frac{x_i-X}{\sigma}+z \right) \left(\mathbbm{1}_{\{   \frac{|x_i-X|}{\sigma} \le d \}} + \mathbbm{1}_{\{   \frac{|x_i-X|}{\sigma} > d \}} \right) \\
& \le d B +  \frac{|x_i-X|}{\sigma} f_Z\left(\frac{x_i-X}{\sigma}+z \right)  \mathbbm{1}_{\{   \frac{|x_i-X|}{\sigma} > d \}} \\
& \le d B + C, \label{eq:bound_for_dc}
    \end{align}
    where $\sup_z f(z) \le B$ and 
    where $C<\infty$ which exists since  $f_Z(u)=o(|u|^{-1})$.

    Now combining the bound in \eqref{eq:bound_for_dc} with the dominated convergence theorem, we arrive at
    \begin{equation}
  \lim_{\sigma \to 0}  \E\left[ \frac{|x_i-X|}{\sigma} f_Z\left(\frac{x_i-X}{\sigma}+z \right)\right]  =0 . \label{eq:limit_top}
    \end{equation}

    
    As a consequence of \eqref{eq:firstBound_on_g}, \eqref{eq:limit_top} and $f_Z(Z)>0$ a.s., we have that $\lim_{\sigma \to 0} |{\cal E}_\sigma| = 0$ a.s., 
    and the claim follows.
\end{IEEEproof}

\subsection{Mixed Random Variables}
In the previous two sections, we have considered the case where the input $X$ was discrete or continuous.  We now consider the case of mixed random variables and assume that $X$ is composed as follows:
\begin{equation}
X =  \mathbbm{1}_{\{U=1\}}  X_1 + \mathbbm{1}_{\{U=2\}} X_2, \label{eq:mixed_input}
\end{equation}
where $X_i$ is a random variable with distribution $\mu_i$, for $i=1,2$, and $U$ is a random variable independent of $X_1,X_2$, taking values on $\{1,2\}$ with probability $\mathbb{P}[U=1]= \alpha \in (0,1)$.
Note that  the distribution of $X$ is 
\begin{equation}
    \mu = \alpha \mu_1 + (1-\alpha) \mu_2.
\end{equation}
Of particular interest in this section is the case when $\mu_1 \perp \mu_2$, which can include the case when $\mu_1$ is discrete and $\mu_2$ is continuous.  

Before proceeding with the general result, we also need to impose regularity conditions on the additive noise. We adopt the same regularity conditions as those used in \cite{wu2011mmse} which are fairly general. 

\begin{definition}
We refer to a random variable $Z$ as a Doob's random variable if its density satisfies the conditions in  \cite[Thm.~4.1]{doob1960relative} (see also Lemma~1 in Technical Appendix~A).
\end{definition}
Loosely speaking, Doob's condition controls the smoothness and the tail behavior of the distribution of $Z$. Many practically relevant random variables, such as the Gaussian, satisfy Doob's conditions.

We begin by proving the following decomposition result.

\begin{theorem}\label{thm:decomp_theorem} Given an input as in \eqref{eq:mixed_input} suppose that 
\begin{itemize}
\item $\mu_1 \perp \mu_2$;  and
\item $Z$ is a Doob's random variable. 
\end{itemize}
and for $u \in \{1,2\}$ define
\begin{align}
Y_u &= X_u + Z, \\
 {\cal E}_{u,\sigma} &:= \frac{X_u - \E[X_u|Y_u]}{\sigma} = -  Z +  \E[Z|Y_u] . 
\end{align}
Then, for every $\alpha \in [0,1]$,
    \begin{equation}
        \lim_{\sigma \to 0} |{\cal E}_\sigma - \mathbbm{1}_{\{U=1\}} {\cal E}_{1,\sigma}-\mathbbm{1}_{\{U=2\}} {\cal E}_{2,\sigma}| = 0  \qquad \text{a.s.}
    \end{equation}
\end{theorem}
\begin{IEEEproof}
   See Technical Appendix~\ref{app:proof_theorem_5}.  
\end{IEEEproof}

Now, as a corollary of Theorem~\ref{thm:decomp_theorem} and previous results, we have the following conclusion. 
\begin{corollary} Suppose that 
\begin{itemize}
\item $X$ in \eqref{eq:mixed_input} be such that  $X_1$  is discrete and $X_2$ has an absolutely continuous distribution; and 
\item $Z$ is a Doob's random variable. 
\end{itemize}
Then, for every $\alpha \in [0,1]$,
\begin{equation}
\lim_{\sigma  \to 0} \mathcal{E}_\sigma = \mathbbm{1}_{\{U=2\}} {\cal E}_{2,\sigma} = \mathbbm{1}_{\{U=2\}}(\mathbb{E}[Z]-Z) \text{ a.s.}
\end{equation}
    
\end{corollary}

We conclude this section by presenting a decomposition theorem for two measures that are absolutely continuous. 

\begin{theorem} \label{thm:decomp_result_abs_measures}
Given an input as in \eqref{eq:mixed_input} suppose that 
\begin{itemize}
\item $\mu_1\ll \mu_2$;  and
\item $Z$ is a Doob's random variable. 
\end{itemize}
Then, for every $\alpha \in [0,1]$,
    \begin{equation}
        \lim_{\sigma \to 0} |{\cal E}_\sigma - \mathbbm{1}_{\{U=1\}} {\cal E}_{1,\sigma}-\mathbbm{1}_{\{U=2\}} {\cal E}_{2,\sigma}| = 0 \qquad \text{a.s.}
    \end{equation}
\end{theorem}
\begin{IEEEproof}
 See Technical Appendix~\ref{app:proof_theorem_6}.
\end{IEEEproof}

As before, Theorem~\ref{thm:decomp_result_abs_measures} is only a decomposition result and limits of ${\cal E}_{2,\sigma}$, if they exists, would have to be found separately. 

\section{Conclusion}
This work departed from the standard moment-based treatment of the estimation error and instead focused on its distributional properties and the pointwise convergence. The main focus was on additive noise channels. The paper derived the structure of the probability density function of the estimation error. Additionally, the pointwise convergence of the estimation error in the low-noise regime was characterized for a fairly general setting.

As a future direction, it would be interesting to characterize similar pointwise limits for the information density and to find an equivalent limit for the information dimension \cite{wu2014information}. It will also be interesting to connect the limits in this work with the pointwise I-MMSE relationship \cite{venkat2012pointwise}. In this work, the conditional mean was used as the estimator, but it would also be interesting to consider other estimators, such as the conditional median \cite{L1Estimaion}.

\begin{appendices}

\section{Limit Theorems}
In this section, we collect limit theorems needed in our analysis. {First, we prove the following technical result:
\begin{lem} \label{lem:bound_on_ratios}
Let $\delta: (0,\infty)\to \mathbb{R}$ and $\delta_1: (0,\infty)\to \mathbb{R}$ be  strictly positive functions with limit $0$ at $0$. Suppose that the following conditions are satisfied
\begin{enumerate}[label=A\arabic*)]
        \item $f_Z(z)$ and $|z| f_Z(z)$ are bounded;
        \item $\liminf_{\sigma\to 0} \inf_{|t|\le \delta(\sigma) } f_Z(z-\frac{t}{\sigma}) >0$ for all $z \in \mathbb{R}$; 
        \item for any $\gamma>0$, 
        \begin{equation}
            K(\sigma, \gamma) := \sup_{|t|\ge \gamma } \left|z- \frac{t}{\sigma} \right| f_Z\left(z-\frac{t}{\sigma}\right) = o(\delta(\sigma))
        \end{equation}
        for all $z \in \mathbb{R}$; for any $\gamma_1>0$, 
        \begin{equation}
            K_1(\sigma, \gamma_1) := \sup_{|t|\ge \gamma_1 }  f_Z\left(z-\frac{t}{\sigma}\right) = o(\delta_1(\sigma))
        \end{equation}
        for all $z \in \mathbb{R}$;
        \item there is a constant $b$ such that 
        \begin{equation}
            K(\sigma, |t|) \le b  \left| z-\frac{t}{\sigma} \right| f_Z\left(z-\frac{t}{\sigma}\right), \quad |t| \in [\delta(\sigma),\gamma],
        \end{equation}
        if $\gamma$ is sufficiently small, for all $z \in \mathbb{R}$; there is a constant $b_1$ such that 
        \begin{equation}
            K_1(\sigma, |t|) \le b_1   f_Z\left(z-\frac{t}{\sigma}\right), \quad |t| \in [\delta_1(\sigma),\gamma_1],
        \end{equation}
        if $\gamma_1$ is sufficiently small, for all $z \in \mathbb{R}$;
        \item if $\gamma>0$, there is a number $\sigma_0$, which can be taken arbitrarily near $0$, for which
        \begin{equation}
         \left| z-\frac{t}{\sigma} \right| f_Z\left(z-\frac{t}{\sigma}\right) \le o(\delta(\sigma)) \left|z -\frac{t}{\sigma_0} \right| f_Z\left(z-\frac{t}{\sigma_0}\right),
         \end{equation}
        for $|t|\ge \gamma$ and for all $z \in \mathbb{R}$.
        Here, $o(\delta(\sigma))$ may depend on $\gamma$ but not on $t$. If $\gamma_1>0$, there is a number $\sigma_{0,1}$, which can be taken arbitrarily near $0$, for which
        \begin{equation}
          f_Z\left(z-\frac{t}{\sigma}\right) \le o(\delta_1(\sigma))  f_Z\left(z-\frac{t}{\sigma_{0,1}}\right),
         \end{equation}
        for $|t|\ge \gamma_1$ and for all $z \in \mathbb{R}$.
        Here, $o(\delta_1(\sigma))$ may depend on $\gamma_1$ but not on $t$.
    \end{enumerate}
    Then, as $\sigma \to 0$, we have
    \begin{equation}
      \frac{|f_{u,\sigma}|}{g_{u,\sigma}}(x+\sigma z) = O(1)  
    \end{equation}
     for $u=\{1,2\}$, all $z\in \mathbb{R}$, and $\mu_u$-a.e. $x$.     
    Furthermore, if $\mu_1 \perp \mu_2$, as $\sigma \to 0$ we have
    \begin{equation}
      \frac{|f_{3-u,\sigma}|}{g_{u,\sigma}}(x+\sigma z) = o(1),\quad \frac{g_{3-u,\sigma}}{g_{u,\sigma}}(x+\sigma z)=o(1), 
    \end{equation}
    for $u \in \{1,2\}$, all $z\in \mathbb{R}$, and $\mu_u$-a.e. $x$. 
\end{lem}
\begin{IEEEproof}
    {Let $\{\sigma_n\}$ for $n \in \mathbb{N}$ be a sequence of positive reals with $\sigma_n \to 0$. Furthermore, let $|\sigma_n z|\le \delta(\sigma_n)$ and denote by ${\cal B}(\mu,\rho)$ the ball of center $\mu$ and radius $\rho$.}
{Then, we have
\begin{align}
    &\frac{|f_{2,\sigma_n}|}{g_{2,\sigma_n}}(x+\sigma_n z) \le \frac{\E\left[\left|\frac{x-X_2}{\sigma_n}+z\right| f_Z\left(\frac{x-X_2}{\sigma_n}+z \right) \right]}{g_{2,\sigma_n}(x+\sigma_n z)} \\
    &\le \underbrace{\frac{\int_{{\cal B}(\sigma_n z, \delta(\sigma_n))}\left|-\frac{t}{\sigma_n}+z\right| f_Z\left(-\frac{t}{\sigma_n}+z \right) \mu_2(x+\mathrm{d}t) }{g_{2,\sigma_n}(x+\sigma_n z)}}_{I_4} \nonumber \\
    &\quad + \underbrace{\frac{\int_{\delta(\sigma_n )}^\gamma K(\sigma_n, t) \mathrm{d}\mu_2(x+{\cal B}(\sigma_n z, t)) }{g_{2,\sigma_n}(x+\sigma_n z)}}_{I_5} \nonumber\\
    &\quad + \underbrace{\frac{\int_\gamma^\infty K(\sigma_n, t) \mathrm{d}\mu_2(x+{\cal B}(\sigma_n z, t)) }{g_{2,\sigma_n}(x+\sigma_n z)}}_{I_6}. \label{eq:def_I4_I5_I6}
\end{align}
We can lower-bound $g_{2,\sigma_n}$ as follows:
\begin{align}
    &g_{2,\sigma_n}(x+\sigma_n z) = \E\left[f_Z\left(\frac{x-X_2}{\sigma_n}+z \right) \right] \\
    &\ge \mu_2(x+{\cal B}(\sigma_n z, \delta(\sigma_n ))) \inf_{t \in {\cal B}(\sigma_n z, \delta(\sigma_n ))} f_Z\left(z-\frac{t}{\sigma_n} \right), \label{eq:lower_g2}
\end{align}
and write the following as $n\to \infty$:
\begin{align}
    \frac{\mu_2(x+{\cal B}(\sigma_n z, \delta(\sigma_n )))}{g_{2,\sigma_n}(x+\sigma_n z)} &= O(1), \label{eq:bigO1} \\
    \frac{\mu_1(x+{\cal B}(\sigma_n z, \delta(\sigma_n )))}{g_{2,\sigma_n}(x+\sigma_n z)} &= o(1), \label{eq:smallo1} \\
    \frac{\delta(\sigma_n )}{g_{2,\sigma_n}(x+\sigma_n z)} &= O(1), \label{eq:bigO2} \\
    \frac{K(\sigma_n, \gamma)}{g_{2,\sigma_n}(x+\sigma_n z)} &= o(1), \label{eq:smallo2}
\end{align}
for $\mu_2$-a.e. $x$, where \eqref{eq:bigO1} follows from \eqref{eq:lower_g2} and condition A2; \eqref{eq:smallo1} follows from \eqref{eq:bigO1} and $\mu_1 \perp \mu_2$; \eqref{eq:bigO2} follows from \eqref{eq:lower_g2} and condition A2; and \eqref{eq:smallo2} follows from \eqref{eq:bigO2} and condition A3.
By using condition A1 and \eqref{eq:bigO1}, we have
\begin{equation}
    I_4 \le \frac{k \mu_2(x+{\cal B}(\sigma_n z, \delta(\sigma_n )))}{g_{2,\sigma_n}(x+\sigma_n z)} = O(1), \label{eq:upper_I4}
\end{equation}
where $k$ is a constant independent of $\sigma_n$ and of $z$. Integrating by parts, we have
\begin{align}
    I_5 &\le K(\sigma_n, \gamma) \frac{\mu_2(x+{\cal B}(\sigma_n z, \gamma))}{g_{2,\sigma_n}(x+\sigma_n z)} \nonumber\\
    &\quad - \frac{\int_{\delta(\sigma_n )}^\gamma \mu_2(x+{\cal B}(\sigma_n z, t)) \mathrm{d}K(\sigma_n, t)}{g_{2,\sigma_n}(x+\sigma_n z)} \label{eq:I_5_bound}.
\end{align}
The first term on the RHS of \eqref{eq:I_5_bound} is $O(1)$, thanks to \eqref{eq:bigO1} and to condition A1 that bounds function $K$. Integrating by parts again, we have
\begin{align}
    I_5 &\le O(1) +  \frac{\int_{\delta(\sigma_n )}^\gamma K(\sigma_n, t) \mathrm{d}\mu_2(x+{\cal B}(\sigma_n z, t)) }{g_{2,\sigma_n}(x+\sigma_n z)} \\
    &\le O(1) + b (I_4+I_5+I_6), \label{eq:backto_f2_g2} 
\end{align}
where $O(1)$ does not depend on the choice of $\gamma$, and \eqref{eq:backto_f2_g2} follows from condition A4 and from inequality \eqref{eq:def_I4_I5_I6}. Hence, we have
\begin{align}
    I_5 &\le O(1) + \frac{b}{1-b} (I_4 + I_6) \\
    &\le O(1) + \frac{b}{1-b} I_6,
\end{align}
where we have used \eqref{eq:upper_I4}.
By using condition A5, we have
\begin{align}
    I_6 &\le o(\delta(\sigma_n )) \frac{\int \left| -\frac{t}{\sigma_0}+z \right| f_Z\left(-\frac{t}{\sigma_0}+z\right) \mu_2(x+\mathrm{d}t)}{g_{2,\sigma_n}(x+\sigma_n z)} \\
    &= \frac{o(\delta(\sigma_n ))}{g_{2,\sigma_n}(x+\sigma_n z)} \label{eq:use_A1_I6} \\
    &= o(1), \label{eq:use_bigO2_I6}
\end{align}
where in \eqref{eq:use_A1_I6} we used condition A1, and in \eqref{eq:use_bigO2_I6} we used \eqref{eq:bigO2}. To sum up, we have proved that, as $n\to\infty$, 
\begin{equation}
    \frac{|f_{2,\sigma_n}|}{g_{2,\sigma_n}}(x+\sigma_n z) =O(1)
\end{equation}
for all $z\in\mathbb{R}$ and $\mu_2$-a.e.~$x$. In a similar fashion, we can prove that, as $n\to\infty$, 
\begin{equation}
    \frac{|f_{1,\sigma_n}|}{g_{1,\sigma_n}}(x+\sigma_n z) =O(1)
\end{equation}
for all $z\in\mathbb{R}$ and $\mu_1$-a.e.~$x$.
}

{Next, consider
\begin{align}
    &\frac{|f_{1,\sigma_n}|}{g_{2,\sigma_n}}(x+\sigma_n z) \le \frac{\E\left[\left|\frac{x-X_1}{\sigma_n}+z\right| f_Z\left(\frac{x-X_1}{\sigma_n}+z \right) \right]}{g_{2,\sigma_n}(x+\sigma_n z)} \\
    &\le \underbrace{\frac{\int_{{\cal B}(\sigma_n z, \delta(\sigma_n))}\left|-\frac{t}{\sigma_n}+z\right| f_Z\left(-\frac{t}{\sigma_n}+z \right) \mu_1(x+\mathrm{d}t) }{g_{2,\sigma_n}(x+\sigma_n z)}}_{I_1} \nonumber \\
    &\quad + \underbrace{\frac{\int_{\delta(\sigma_n )}^\gamma K(\sigma_n, t) \mathrm{d}\mu_1(x+{\cal B}(\sigma_n z, t)) }{g_{2,\sigma_n}(x+\sigma_n z)}}_{I_2} \nonumber\\
    &\quad + \underbrace{\frac{\int_\gamma^\infty K(\sigma_n, t) \mathrm{d}\mu_1(x+{\cal B}(\sigma_n z, t)) }{g_{2,\sigma_n}(x+\sigma_n z)}}_{I_3}.
\end{align}
 By using condition A1 and \eqref{eq:smallo1}, we have
\begin{equation}
    I_1 \le \frac{k \mu_1(x+{\cal B}(\sigma_n z, \delta(\sigma_n )))}{g_{2,\sigma_n}(x+\sigma_n z)} = o(1),
\end{equation}
where $k$ is a constant independent of $\sigma_n$ and of $z$. Integrating by parts, we have
\begin{align}
    I_2 &\le K(\sigma_n, \gamma) \frac{\mu_1(x+{\cal B}(\sigma_n z, \gamma))}{g_{2,\sigma_n}(x+\sigma_n z)}  \nonumber\\
    &\quad - \frac{\int_{\delta(\sigma_n )}^\gamma \mu_1(x+{\cal B}(\sigma_n z, t)) \mathrm{d}K(\sigma_n, t)}{g_{2,\sigma_n}(x+\sigma_n z)} \label{eq:I_2_bound}.
\end{align}
The first term on the RHS of \eqref{eq:I_2_bound} is $o(1)$, thanks to \eqref{eq:smallo1}. If $\varepsilon>0$, there is, since $\mu_1 \perp \mu_2$, a value of $\gamma$ so small that
\begin{equation}
    \mu_1(x+{\cal B}(\sigma_n z, t)) \le \varepsilon \mu_2(x+{\cal B}(\sigma_n z, t)), \qquad t \in [\delta(\sigma_n ), \gamma].
\end{equation}
With this value of $\gamma$, we find that the RHS of \eqref{eq:I_2_bound} is at most
\begin{equation}
    o(1)-\varepsilon \frac{\int_{\delta(\sigma_n )}^\gamma \mu_2(x+{\cal B}(\sigma_n z, t)) \mathrm{d}K(\sigma_n, t)}{g_{2,\sigma_n}(x+\sigma_n z)}
\end{equation}
and integrating by parts again, we have
\begin{align}
    I_2 &\le o(1) + \varepsilon O(1) + \varepsilon \frac{\int_{\delta(\sigma_n )}^\gamma K(\sigma_n, t) \mathrm{d}\mu_2(x+{\cal B}(\sigma_n z, t)) }{g_{2,\sigma_n}(x+\sigma_n z)} \\
    &\le o(1) + \varepsilon O(1) + b\varepsilon(I_4+I_5+I_6) \label{eq:use_I4_I5_I6} \\
    &\le o(1) + \epsilon O(1),
\end{align}
where $O(1)$ does not depend on the choice of $\gamma$; \eqref{eq:use_I4_I5_I6} follows from conditions A4 and from inequality \eqref{eq:def_I4_I5_I6}; and the last step follows from $I_4+I_5+I_6 \le O(1)$. Therefore, $I_2$ can be made arbitrarily small by choosing $\gamma$ sufficiently small. By using condition A5, we have
\begin{align}
    I_3 &\le o(\delta(\sigma_n )) \frac{\int \left| -\frac{t}{\sigma_0}+z \right| f_Z\left(-\frac{t}{\sigma_0}+z\right) \mu_1(x+\mathrm{d}t)}{g_{2,\sigma_n}(x+\sigma_n z)} \\
    &= \frac{o(\delta(\sigma_n ))}{g_{2,\sigma_n}(x+\sigma_n z)} \label{eq:use_A1} \\
    &= o(1), \label{eq:use_bigO2}
\end{align}
where in \eqref{eq:use_A1} we used condition A1, and in \eqref{eq:use_bigO2} we used \eqref{eq:bigO2}. To sum up, we have proved that, as $n\to\infty$, 
\begin{equation}
    \frac{|f_{1,\sigma_n}|}{g_{2,\sigma_n}}(x+\sigma_n z) =o(1)
\end{equation}
for all $z\in\mathbb{R}$ and $\mu_2$-a.e. $x$. In a similar fashion, we can prove that, as $n\to\infty$, 
\begin{equation}
    \frac{|f_{2,\sigma_n}|}{g_{1,\sigma_n}}(x+\sigma_n z) =o(1)
\end{equation}
for all $z\in\mathbb{R}$ and $\mu_1$-a.e. $x$.

Finally, if $\mu_1 \perp \mu_2$, we can apply \cite[Lemma~5]{wu2011mmse} to state that 
\begin{equation}
      \frac{g_{3-u,\sigma}}{g_{u,\sigma}}(x+\sigma z)=o(1), 
    \end{equation}
    for $u=\{1,2\}$, all $z\in \mathbb{R}$, and $\mu_u$-a.e. $x$. 
}
\end{IEEEproof}

\section{Proof of Theorem \ref{thm:decomp_theorem}}\label{app:proof_theorem_5}

The case of $\alpha \in \{0,1\}$ has been treated in previous sections and, here, we focus on $\alpha \in (0,1)$.  Next, for $u \in \{1,2\}$ define 
\begin{align}
    g_{u,\sigma}(y) &= \E\left[f_Z\left(\frac{y-X_u}{\sigma} \right) \right], \label{eq:def_g} \\
    f_{u,\sigma}(y) &= \E\left[\frac{y-X_u}{\sigma}f_Z\left(\frac{y-X_u}{\sigma} \right) \right] \label{eq:def_f}
\end{align}
and
\begin{align}
    g_\sigma &= \alpha_1 g_{1,\sigma} + \alpha_2 g_{2,\sigma}, \\
    f_\sigma &= \alpha_1 f_{1,\sigma} + \alpha_2 f_{2,\sigma},
\end{align}
where $\alpha_1=\alpha $ and $\alpha_2 = 1-\alpha$. 
Then, the densities of $Y_u = X_u + \sigma Z$ and $Y = X + \sigma Z$ are respectively given by
\begin{align}
    q_{u,\sigma}(y) &= \frac{1}{\sigma} g_{u,\sigma}(y), \\
    q_{\sigma}(y) &= \frac{1}{\sigma} g_{\sigma}(y). 
\end{align}

Next, suppose that  $ U = u \in  \{1,2\}$
\begin{align}\label{eq:diff_calE}
    |{\cal E}_{\sigma}-{\cal E}_{u,\sigma}| = |\E[Z|Y]-\E[Z|Y,U=u]|.
\end{align}

Let us compute
\begin{align}
    &|\E[Z|Y=y]-\E[Z|Y=y,U=u]| \nonumber \\
    &= \left|\frac{f_\sigma}{g_\sigma}-\frac{f_{u,\sigma}}{g_{u,\sigma}} \right| \circ y \\
    &= \frac{|f_\sigma g_{u,\sigma}-f_{u,\sigma} g_\sigma|}{g_\sigma g_{u,\sigma}} \circ y \\
    &= \frac{|(\alpha_1 f_{1,\sigma} + \alpha_2 f_{2,\sigma}) g_{u,\sigma}-f_{u,\sigma} (\alpha_1 g_{1,\sigma} + \alpha_2 g_{2,\sigma})|}{g_\sigma g_{u,\sigma}} \circ y \\
    &=\left\{
    \begin{array}{cc}
        \alpha_2 \frac{| f_{1,\sigma}  g_{2,\sigma}-f_{2,\sigma}  g_{1,\sigma} |}{g_\sigma g_{1,\sigma}} \circ y & u=1 \\
        \alpha_1 \frac{| f_{1,\sigma}  g_{2,\sigma}-f_{2,\sigma}  g_{1,\sigma} |}{g_\sigma g_{2,\sigma}} \circ y & u=2
    \end{array}
    \right. \\
    &=\left\{
    \begin{array}{cc}
        \alpha_2 \left(\left|\frac{ f_{1,\sigma}  }{g_{1,\sigma}}-\frac{ f_{2,\sigma}  }{g_{2,\sigma}}\right| \frac{g_{2,\sigma}}{g_\sigma}\right) \circ y & u=1 \\
        \alpha_1 \left(\left|\frac{ f_{1,\sigma}  }{g_{1,\sigma}}-\frac{ f_{2,\sigma}  }{g_{2,\sigma}}\right| \frac{g_{1,\sigma}}{g_\sigma}\right) \circ y & u=2.
    \end{array}
    \right. \label{eq:diff_E}
\end{align}

Next, consider
\begin{align}
   &\left(\left|\frac{ f_{1,\sigma}  }{g_{1,\sigma}}-\frac{ f_{2,\sigma}  }{g_{2,\sigma}}\right| \frac{g_{3-u,\sigma}}{g_\sigma}\right) \circ (x+\sigma z) \nonumber \\
     &\le    \left( \left( \frac{|f_{1,\sigma}|}{g_{1,\sigma}}+\frac{|f_{2,\sigma}|}{g_{2,\sigma}} \right) \frac{g_{3-u,\sigma}}{\alpha_u g_{u,\sigma}} \right) \circ (x+\sigma z).  \label{eq:upper_diff_EX_given_Y}
\end{align} 
To conclude the proof,  we show that  for $u \in \{1,2 \}$
\begin{align}
 &  \frac{|f_{u,\sigma}|}{g_{u,\sigma}}  \circ (x+\sigma z) = O(1), \quad \sigma \to 0, \label{eq:boundedness:diff_cal} \\ 
&   \frac{g_{3-u,\sigma}}{ g_{u,\sigma}}  \circ (x+\sigma z) = o(1), \quad \sigma \to 0,  \label{eq:use_lemma5}
\end{align}
for every $z \in \mathbb{R}$ and $\mu_u$-a.e. $x$; where  \eqref{eq:boundedness:diff_cal} follows from Lemma~\ref{lem:bound_on_ratios}; and \eqref{eq:use_lemma5} follows from 
$\mu_1 \perp \mu_2$ and from Lemma~\ref{lem:bound_on_ratios}.

Putting together \eqref{eq:diff_calE}, \eqref{eq:diff_E}, and \eqref{eq:use_lemma5}, we have
\begin{equation}
    |{\cal E}_{\sigma}-{\cal E}_{u,\sigma}| = o(1), \qquad \sigma \to 0,
\end{equation}
for all $z \in \mathbb{R}$, $u=\{1,2\}$, and $\mu_u$-a.e. $x$.


\section{Proof of Theorem~\ref{thm:decomp_result_abs_measures}}
\label{app:proof_theorem_6}
Consider $\mu_1 \ll \mu_2$. We first need to introduce the following technical result, which is a slightly modified version of \cite[Lemma~6]{wu2011mmse}.
\begin{lem}\label{lem:differentiation_thm} Let $\nu$ be a probability measure and $g\in L^1_{\text{loc}}(\mathbb{R},\nu)$. Then,
    under the conditions of Lemma \ref{lem:bound_on_ratios}, 
    \begin{equation}
        \lim_{\sigma\to 0} \frac{\int |g(t)-g(x)| \left|\frac{t-x}{\sigma}+z\right| f_Z\left(\frac{t-x}{\sigma}+z \right) \nu(\mathrm{d}t) }{\int  \left|\frac{t-x}{\sigma}+z\right| f_Z\left(\frac{t-x}{\sigma} +z\right) \nu(\mathrm{d}t)}=0
    \end{equation}
    holds for $\nu$-a.e. $x$ and every $z \in \mathbb{R}$.
\end{lem}

Now, let $h := \frac{\mathrm{d}\mu_1}{\mathrm{d}\mu_2}$
    be the Radon-Nikodym derivative of $\mu_1$ with respect to $\mu_2$, where $h:\mathbb{R}\to \mathbb{R}^{+}$ satisfies $\int h \mathrm{d}\mu_2 = 1$. Define the set
    \begin{equation}
        {\cal F} = \{x: \: h(x)>0\}.
    \end{equation}
    From \eqref{eq:diff_E}, for $u=\{1,2\}$ we have
    \begin{align}
    &|{\cal E}_{\sigma}-{\cal E}_{u,\sigma}| = \alpha_{3-u} \left(\left|\frac{ f_{u,\sigma}  }{g_{u,\sigma}}-\frac{ f_{3-u,\sigma}  }{g_{3-u,\sigma}}\right| \frac{g_{3-u,\sigma}}{g_\sigma}\right) \circ y \label{eq:Esigma_minus_Esigmau}\\
    &\le  \alpha_{3-u} \left(\left(\frac{ |f_{1,\sigma}|  }{g_{1,\sigma}}+\frac{ |f_{2,\sigma}|  }{g_{2,\sigma}}\right) \frac{g_{3-u,\sigma}}{g_\sigma}\right) \circ y.
\end{align}
Now, by using Lemma~\ref{lem:bound_on_ratios}, as $\sigma \to 0$ we have
\begin{equation}
    \left(\frac{ |f_{1,\sigma}|  }{g_{1,\sigma}}+\frac{ |f_{2,\sigma}|  }{g_{2,\sigma}}\right) \circ (x+\sigma z) = O(1)
\end{equation}
for every $z \in \mathbb{R}$ and $\mu_2$-a.e. $x$.

For $x \in {\cal F}^c$, we necessarily have $u=2$ and
\begin{equation}
    \frac{g_{1,\sigma}}{g_\sigma}(x+\sigma z) \le \frac{g_{1,\sigma}}{g_{2,\sigma}}(x+\sigma z) = o(1), \qquad \sigma \to 0,
\end{equation}
for $\mu_2$-a.e. $x$ and every $z$, where the last step is thanks to \cite[Lemma 5]{wu2011mmse}.  
As a consequence, for $x \in {\cal F}^c$ we can write that
\begin{align}
   &|({\cal E}_\sigma - \mathbbm{1}_{\{u=1\}} {\cal E}_{1,\sigma}-\mathbbm{1}_{\{u=2\}} {\cal E}_{2,\sigma})(x+\sigma z)| \nonumber\\
   &=  |({\cal E}_{\sigma}-{\cal E}_{2,\sigma})(x+\sigma z)| = o(1), \qquad \sigma\to 0,
\end{align}
holds for $\mu_2$-a.e. $x$ and every $z\in\mathbb{R}$. 

Define
\begin{equation}
    v(x+\sigma z) := \E\left[\left|\frac{x-X_2}{\sigma}+z \right| f_Z\left(\frac{x-X_2}{\sigma}+z \right) \right].
\end{equation}
For $x \in {\cal F}$, write 
\begin{align}
    &|(f_{1,\sigma}g_{2,\sigma}-f_{2,\sigma}g_{1,\sigma})(x+\sigma z)| \nonumber\\
    &= \left|\int \left(\frac{x-x_1}{\sigma}+z \right) f_Z\left(\frac{x-x_1}{\sigma}+z\right) h(x_1) \mu_2(\mathrm{d}x_1)\right.\nonumber\\
    &\quad\quad \int f_Z\left(\frac{x-x_2}{\sigma}+z\right)  \mu_2(\mathrm{d}x_2)  \nonumber \\
    & \quad  -\int \left(\frac{x-x_1}{\sigma}+z \right) f_Z\left(\frac{x-x_1}{\sigma}+z\right)  \mu_2(\mathrm{d}x_1) \nonumber\\
    &\quad\quad  \left.\int f_Z\left(\frac{x-x_2}{\sigma}+z\right)  h(x_2)\mu_2(\mathrm{d}x_2)\right| \\& \le  \int \int |h(x_1)-h(x_2)| \left|\frac{x-x_1}{\sigma}+z \right| f_Z\left(\frac{x-x_1}{\sigma}+z\right) \nonumber\\
    &\quad f_Z\left(\frac{x-x_2}{\sigma}+z\right) \mu_2(\mathrm{d}x_1)\mu_2(\mathrm{d}x_2) \\
    &\le \int \int |h(x_1)-h(x)| \left|\frac{x-x_1}{\sigma}+z \right| f_Z\left(\frac{x-x_1}{\sigma}+z\right) \nonumber\\
    &\quad\quad  f_Z\left(\frac{x-x_2}{\sigma}+z\right) \mu_2(\mathrm{d}x_1)\mu_2(\mathrm{d}x_2)  \nonumber\\
    &\quad + \int \int |h(x)-h(x_2)| \left|\frac{x-x_1}{\sigma}+z \right| f_Z\left(\frac{x-x_1}{\sigma}+z\right) \nonumber\\
    &\quad \quad f_Z\left(\frac{x-x_2}{\sigma}+z\right) \mu_2(\mathrm{d}x_1)\mu_2(\mathrm{d}x_2) \\
    &= g_{2,\sigma}(x+\sigma z) \nonumber\\
    &\quad \int |h(x_1)-h(x)| \left|\frac{x-x_1}{\sigma}+z \right| f_Z\left(\frac{x-x_1}{\sigma}+z\right) \mu_2(\mathrm{d}x_1) \nonumber\\
    &\quad + \int \left|\frac{x-x_1}{\sigma}+z \right| f_Z\left(\frac{x-x_1}{\sigma}+z\right)\mu_2(\mathrm{d}x_1) \nonumber\\
    &\quad \int |h(x)-h(x_2)| f_Z\left(\frac{x-x_2}{\sigma}+z\right) \mu_2(\mathrm{d}x_2) \\
    &= g_{2,\sigma}(x+\sigma z) \cdot o(v(x+\sigma z)) \nonumber\\
    &\quad +v(x+\sigma z)\cdot o(g_{2,\sigma}(x+\sigma z)),\label{eq:diff_f1g2_f2g1}
\end{align}
as $\sigma \to 0$, 
for $\mu_2$-a.e. $x$ and every $z \in \mathbb{R}$, which follows from applying Lemma \ref{lem:differentiation_thm} to $h \in L^1(\mathbb{R}, \mu_2)$.

 By plugging \eqref{eq:diff_f1g2_f2g1} into \eqref{eq:Esigma_minus_Esigmau}, we have
\begin{align}
    |{\cal E}_{\sigma}-{\cal E}_{u,\sigma}| &\le \alpha_{3-u} \frac{g_{2,\sigma}(x+\sigma z) \cdot o(v(x+\sigma z))}{g_{u,\sigma}(x+\sigma z) g_\sigma(x+\sigma z)} \nonumber \\
    &\quad +\alpha_{3-u} \frac{v(x+\sigma z)\cdot o(g_{2,\sigma}(x+\sigma z))}{g_{u,\sigma}(x+\sigma z) g_\sigma(x+\sigma z)} \\
    &= \alpha_{3-u} \frac{g_{2,\sigma}(x+\sigma z) v(x+\sigma z) }{g_{u,\sigma}(x+\sigma z) g_\sigma(x+\sigma z)} \nonumber\\
    &\quad \cdot \left( \frac{o(v(x+\sigma z))}{v(x+\sigma z)}+\frac{o(g_{2,\sigma}(x+\sigma z))}{g_{2,\sigma}(x+\sigma z)}\right)
\end{align}
for $u \in \{1,2\}$.
 Next, by \cite[Lemma 5]{wu2011mmse} and by Lemma \ref{lem:bound_on_ratios}, we have
\begin{align}
     \frac{g_{2,\sigma}(x+\sigma z)  v(x+\sigma z)}{g_{u,\sigma}(x+\sigma z) g_\sigma(x+\sigma z)} &\le  \frac{g_{2,\sigma}(x+\sigma z) v(x+\sigma z)}{g_{u,\sigma}(x+\sigma z)\cdot  \alpha_2 g_{2,\sigma}(x+\sigma z)} \\
    &=\left\{
    \begin{array}{cc}
        \frac{1}{\alpha_2 h(x) }O(1) & u=1 \\
        \frac{1}{ \alpha_2}O(1)  & u=2
    \end{array}
     \right. \\
     &=O(1), \quad \sigma\to 0,
\end{align}
for $\mu_2$-a.e. $x$. Putting everything together, we have
\begin{equation}
        \lim_{\sigma \to 0} |{\cal E}_\sigma - \mathbbm{1}_{\{U=1\}} {\cal E}_{1,\sigma}-\mathbbm{1}_{\{U=2\}} {\cal E}_{2,\sigma}| = 0, \qquad \text{a.s.}
    \end{equation}

}






\end{appendices}

\bibliographystyle{ieeetr}
\bibliography{biblio.bib}

\end{document}